# Automatic Wrappers for Large Scale Web Extraction

Nilesh Dalvi
Yahoo! Research
Santa Clara, CA, USA
ndalvi@yahoo-inc.com

Ravi Kumar
Yahoo! Research
Santa Clara, CA, USA
ravikuma@yahoo-inc.com

Mohamed Soliman*
U. of Waterloo
Ontario, Canada
m2ali@cs.uwaterloo.edu

## ABSTRACT

We present a generic framework to make wrapper induction algorithms tolerant to noise in the training data. This enables us to learn wrappers in a completely unsupervised manner from automatically and cheaply obtained noisy training data, e.g., using dictionaries and regular expressions. By removing the site-level supervision that wrapper-based techniques require, we are able to perform information extraction at web-scale, with accuracy unattained with existing unsupervised extraction techniques. Our system is used in production at Yahoo! and powers live applications.

## 1. INTRODUCTION

Several websites use scripts to generate highly structured HTML: this includes shopping sites, entertainment sites, academic repositories, library catalogs, and virtually any website that uses forms to fetch content from a database and serve it to the users. The structural similarity of script-generated webpages is a powerful feature that allows information extraction systems to use simple rules to effectively extract information from these websites. For example, consider the dealer locator form at albanyindustries.com/dealers.asp. Given any zipcode, this form generates a page that lists the dealers in the zipcode. An HTML snippet of such a webpage for the zipcode 38652 is shown in Figure 1. (Figure 5 in Appendix shows the rendered webpage.) When viewed as an XML document tree, the following simple xpath extracts all the dealer names from this webpage:

$$//\texttt{div}[@\texttt{class} =' \texttt{dealerlinks}']/\texttt{tr}/\texttt{td}/\texttt{u}/\texttt{text}()$$

This rule says that all the dealer names can be found under u tag, inside a td tag, inside a tr tag, which is within a div whose class is dealerlinks.

The rule above not only works on this particular webpage, it works on *any* page generated by this form. Such a rule is called a *wrapper*, and the problem of inducing wrappers from labeled examples has been extensively studied [15, 13, 11, 19, 18, 1]. For instance, such a rule can easily be learned if a few business names in few such pages are labeled, e.g., if labels 1 and 2 are specified

*Most of the work was done while the author was at Yahoo! Research.

```
<div class='dealer links'>
  <tr><td>
        <u>PORTER FURNITURE</u><br>   (1)
        201 HWY.30 West<br>
        NEW ALBANY, MS 38652
     </td>
     …
  </tr>
  <tr><td>
        <u>WOODLAND FURNITURE</u><br>   (2)
        123 Main St.<br>
        WOODLAND, MS 3977   (3)
     </td>
     …
  </tr>
</div>
```

**Figure 1: A sample HTML snippet.**

in Figure 1. Wrappers provide an effective mechanism to extract information for a given website, and can often be learned using a very small number of labeled examples.

**Wrappers and annotations.** Wrappers, however, only work at site-level, which traditionally has put a fundamental limitation on their use for web-scale extraction. For example, consider the following task: extract business listings from all the store locator pages on the Web. (Compiling such a database can be immensely useful and can enable powerful Web applications.) For each store locator page, we can use automatic form-filling techniques [16] to easily generate a large collection of HTML pages, each corresponding to store listing in a particular location. While pages for each store locator, being script-generated, have similar structure that is amenable to rule-based extraction, a human still needs to label a few sample pages from every single store locator, making wrappers *infeasible* for this extraction task.

Fortunately, for several domains, it is relatively easy to automatically obtain a decent set of annotations. E.g., for business listings, if we compile a small dictionary of popular business names (Office Depot, BestBuy, etc.), we can annotate these names when they appear among the store locator listings. Alternatively, we can identify certain names containing words like "*.Inc*" and "*Shop*" to most likely be business names.

Such an automated annotation process will inherently be *noisy*, and some fraction of labels will be incorrect. Traditional wrapper induction methods are not designed to handle noise in the training data, and often a single bad annotation can result in an incorrect extraction rule. For example, in Figure 1, suppose we are given the incorrect label 3 along with the two correct labels 1 and 2. Then, in order to accommodate all the labels, the learned rule might be

$$//\texttt{div}[@\texttt{class} =' \texttt{dealerlinks}']/\texttt{tr}/\texttt{td}//\texttt{text}(),$$

which extracts all the text nodes under the td tag. Thus, a single incorrect label can grossly over-generalize the extraction rule.

The focus of our work is to enable noise-tolerant wrapper induction, allowing us to learn wrappers from automatically and cheaply obtained noisy training data. By removing the site-level supervision that wrapper-based techniques require, we are able to perform information extraction at web-scale with high accuracy.[1].

**Our contributions.** We present a generic framework for making supervised wrapper induction noise-tolerant. Given any wrapper inductor that satisfies mild technical conditions, the framework shows how to use it as a blackbox when the labels of the training data are noisy. The framework is based on a natural modeling of the web publication process using domain-specific knowledge and a probabilistic modeling of the noisy labeling process.

The main idea is to *generate-and-test*: given a set of labels that are noisy, we enumerate all possible wrappers generated by subsets of the labels. Each such wrapper is ranked according to its quality, which depends on its likelihood of generating the (noisy) labels according to our annotation model and its likelihood of being a good wrapper as guided by our web publication model. The intuition is that if the label noise is not too excessive, then one of the generated wrappers will be trained on sufficiently many noise-free labels and our careful definition of quality will cause it to be ranked high. En route, we obtain provably efficient algorithms for the enumeration problem that may be of independent interest.

We demonstrate the generality of our framework by applying it to two wrapper induction algorithms from the literature, namely, the WIEN system from Kushmerick et al. [15, 14] and the xpath-based method from Dalvi et al. [6]. Using these, we perform an extensive set of experiments covering over 300 websites over multiple domains, and are able to achieve very high precision and recall. Our system is used in production in Yahoo! for large-scale high-quality information extraction from the Web, powering live search applications.

# 2. MODEL AND PROBLEM

We formally describe here our model for web publication and annotations, and the problem of extracting structured information from webpages.

**Overview.** Our objective is to learn, given a website, a wrapper that can extract the information of interest from the website. We want to use automatic annotators to obtain our training data. Thus, we want to model the possibility that the annotations have errors. We call this the *annotation* model. Further, in the presence of errors, we want to use the background knowledge about websites to distinguish between good and bad wrappers. We call this the *web publication model*. In Section 2.1, we describe the two models, and then in Section 2.2, we formally define our extraction problem in terms of the two models.

## 2.1 Publication and annotation models

**Web publication model.** We now state a model for generating a set of webpages in a given domain of interest. First, we pick a schema $S$ over a set $\mathcal{T}$ of types. Second, we pick a rendering script that takes elements of the schema $S$ and renders them into HTML documents. Third, we pick a set of elements over $S$ and apply the rendering script over them to obtain a set $\mathcal{H}$ of HTML webpages.

[1]It is important to note that our aim is *not* to propose a new wrapper induction language/algorithm but rather show how to incorporate an existing wrapper induction system. There are several languages proposed in the literature for expressing wrappers, e.g., prefix-suffix pairs [15, 14], finite-state automaton [17] and xpaths [6, 1, 18], and various algorithms designed for learning wrappers.

We take the domain of *business listings* as our running example. Consider a set of webpages from the website of a brand, each listing its authorized stores in a particular city. The set $\mathcal{T}$ of types is $\{$*name*, *address*, *phone*$\}$ and the schema is given by $S = (name, address, phone)^*$, denoting a list of tuples, each having *name*, *address*, and *phone*. The rendering script might display it as a table, with a `td` tag enclosing each tuple and a `tr` tag enclosing each attribute, and additionally a `<b>` tag enclosing each *name*. The same script is used to generate webpages for displaying the list of stores for each city.

**Annotation model.** Let $\mathcal{H}$ be a set of webpages. Each webpage contains a set of text elements and a set of tags. We represent the entire $\mathcal{H}$ using a single vector $\hat{A} = \langle A_1, \ldots, A_n \rangle$, where each $A_i$ is either a text element or an HTML tag in some page in $\mathcal{H}$. Each $A_i$ either belongs to one of the types in $\mathcal{T}$ or is part of the script.

We assume that we have a set of annotators, each corresponding to a type, where each annotator labels a subset of the text elements. We use $L$ to denote the set of all annotations, where each annotation is of the form $(A_i, \tau)$ where $A_i$ is a node and $\tau \in \mathcal{T}$ is a type.

We assume that the annotators are noisy. So an annotator for a type $\tau$ might miss some nodes of type $\tau$ and might incorrectly label some nodes that are not of type $\tau$. We characterize annotators by their *precision* $p_\tau$ and *recall* $r_\tau$. For example, consider the dictionary-based business name annotator described in Section 1. This will have a low recall, depending on the size of the dictionary, and will have a high, but not perfect precision, since it can occasionally match words outside the list that look like business names.

## 2.2 The extraction problem

Given a set of webpages represented as a vector $\hat{A}$, we are interested in finding the data items that generated the webpages. Formally, we want to deduce the most likely type for $A_i$'s of interest, i.e., output a partial mapping $X : \hat{A} \to \mathcal{T}$. We are given a set of annotations $L$ over $\hat{A}$ and we wish to maximize $\mathbf{P}(X \mid L)$. We can rewrite this as $\mathbf{P}(L \mid X)\mathbf{P}(X)/\mathbf{P}(L)$. Since $L$ is a constant, we can formulate the extraction problem as finding $X$ given by

$$\arg\max_X \mathbf{P}(L \mid X)\mathbf{P}(X). \tag{1}$$

The first term $\mathbf{P}(L \mid X)$ models the errors in the annotation process. It gives the probability of obtaining the label set $L$ on $\hat{A}$ if the true labels were given by $X$. The second term $\mathbf{P}(X)$ models the prior probability of obtaining $X$ as given by the web publication process, e.g., if $X$ has a good repeating structure, then its probability should be higher. Notice that the extraction problem as formulated in Equation (1) is not yet fully-specified since we still have not specified the probability distributions $\mathbf{P}(L \mid X)$ and $\mathbf{P}(X)$. We give the formal definitions in Section 3.

In general, we are only interested in extracting a subset of types, and not the full grammar that generated the webpage. E.g. if a website has a list of businesses with *name*, *address*, a *link* to map, and a *flag* for featured businesses, we might only be interested in extracting the names and addresses. In this case, we only want to construct a partial mapping $X$ that maps nodes of types *name* and *address* and leaving other nodes unmapped without caring if these other nodes map to *link* or *flag* or are part of the script.

For the sake of presentation, we will present all our results for a simplified version of the problem, which we call the *single-type extraction*, in the main body of the paper. In this problem, we are only interested in extracting a single type $\tau$, and we also assume that we only have annotations of type $\tau$. Thus, we can assume $L$ to be simply a set of nodes (implicitly, of type $\tau$). Similarly, we will assume $X$ to denote a set of nodes, rather than a mapping from nodes to types. In our running example of business listings, we will consider the problem of extracting the business names.

Although we only discuss single-type extraction, our techniques are general and it is straightforward to apply them to the multi-type extraction problem; for completeness, Appendix A contains the details, including experimentation with multi-type extraction.

**Our formulation vs. existing work.** Equation (1) ties together two existing lines of work on extracting structured data from the web. The first line of work is on grammar learning [2, 5], which focus on the web publication process, and try to infer the grammar that generated the given set of webpages. This can be viewed as optimizing the $\mathbf{P}(X)$ term. Union-free regular expressions and its variants have been proposed as a good abstraction for describing the structure of websites. This line of work tries to find the $X$ that can be explained by the smallest grammar, in a suitably chosen language.

The second line of work [14, 15, 6] uses rule induction to learn wrappers from a small number of labeled examples. Several languages have been proposed for expressing the rules that vary in their expressibility and efficiency. This line of work can be viewed as optimizing the $\mathbf{P}(L \mid X)$ term in Equation (1). However, they assume that the annotation process is noise-free. Thus, $\mathbf{P}(L \mid X)$ is 1 if $X$ is consistent with $L$, and 0 otherwise. The problem in this setting is to find a rule, in a suitably chosen wrapper language, that results in an $X$ consistent with the labels.

# 3. OUR APPROACH

We assume that we have a rule induction system $\phi$ that learns a rule in a specific language from a set of noise-free labeled examples $L$, e.g., Wien [15] or XPath learner [6]. Our objective is to make $\phi$ noise-tolerant.

EXAMPLE 1. *To illustrate our approach, we consider a simple hypothetical wrapper inductor that we call* TABLE. *This inductor works on a table, like the one shown below.*

| | name | | | |
|---|---|---|---|---|
| *business*$_1$ | ($n_1$) | $a_1$ | $z_1$ | $p_1$ |
| *business*$_2$ | ($n_2$) | $a_2$ | $z_2$ | $p_2$ |
| *business*$_3$ | $n_3$ | $a_3$ | $z_3$ | $p_3$ |
| *business*$_4$ | ($n_4$) | ($a_4$) | $z_4$ | $p_4$ |
| *business*$_5$ | $n_5$ | $a_5$ | ($z_5$) | $p_5$ |

*If $L$ consists of a single label, e.g., $\{n_1\}$, then* TABLE *learns a rule that returns just the label itself. If $L$ consists of labels all from the same row (or column), e.g., $\{n_4, a_4\}$, then* TABLE *generalizes it to the entire row (or column). If $L$ consists of labels that spans at least two rows and columns, e.g., $\{a_4, z_5\}$, then* TABLE *generalizes it to the entire table.*

*We will use the table above as our running example. In this table, each row contains a business listing and the first column of each row contains the business name. We have five labels as shown in the table above. However, two of them are wrong. We want to learn a wrapper to extract all the names, which, when expressed as a rule in* TABLE, *is simply the first column.*

Given a wrapper inductor $\phi$, which in the example above is TABLE, we want to find the wrapper, among the space of all possible wrappers, that maximizes the probability as given in Equation (1).

**Enumeration.** The first step is to enumerate the space of wrappers. In Example 1, since $L$ contains 5 labels and we do not know which of them are correct, each of the 32 possible subsets of $L$, when given as input to $\phi$, can potentially give us the correct wrapper. In principle, we want to consider wrappers from this space, and rank the resulting wrappers using Equation (1) to pick the best wrapper.

This brings us to the first issue: *efficient enumeration* of the wrapper space. In Example 1, the 32 subsets of $L$ only result in 8 unique wrappers: each of the five individual cells, the first column, the fourth row and the entire table. Denoting row $i$ by $R_i$, column $i$ by $C_i$, and the entire table by $T$, we have the following 8 wrappers:

$$\{\{n_1\}, \{n_2\}, \{n_4\}, \{a_4\}, \{z_5\}, C_1, R_4, T\}. \quad (2)$$

In general, if given all possible $n^2$ labels on an $n \times n$ table, the $2^{n^2}$ subsets will result in only $n^2 + 2n + 1$ unique wrappers for the TABLE wrapper inductor.

This motivates the following *enumeration problem*: given a wrapper inductor $\phi$ and a set of labels $L$, output the set of unique wrappers that the subsets of $L$ generate, without calling $\phi$ on each of the $2^L$ subsets. We present an efficient solution to the enumeration problem in Section 4.

**Ranking.** The second issue at hand is the *ranking* of the resulting wrappers. We want to use Equation (1) to rank wrappers. We again illustrate the ideas behind ranking using the table in Example 1. Suppose we have three wrappers that we want to rank: $w_1$, which consists of the first column, $w_2$, which consists of the first two columns, and $w_3$, which is the entire table. Note that $w_1$ and $w_3$ are expressible in the TABLE wrapper language, while $w_2$ is a hypothetical wrapper included only for the sake of illustration.

Let $X_1$ denote the set of nodes in the first column, $X_2$ denote the set of nodes in the first two columns, and $X_3$ denote the set of all nodes in the table. Thus, the score of $w_i$ is $\mathbf{P}(L \mid X_i)\mathbf{P}(X_i)$.

Let us look at the first component of the score, $\mathbf{P}(L \mid X_i)$, which corresponds to the labeling errors. $\mathbf{P}(L \mid X_1)$ is the probability that the labeling process results in 2 mistakes and 3 correct labels. $\mathbf{P}(L \mid X_2)$ is the probability that there is 1 mistake and 4 correct labels, while $\mathbf{P}(L \mid X_3)$ is the probability that there are no mistakes. When the probability of errors is low, we expect

$$\mathbf{P}(L \mid X_1) < \mathbf{P}(L \mid X_2) < \mathbf{P}(L \mid X_3).$$

Note that the recall of the annotator also plays a role in the above equation. E.g., $\mathbf{P}(L \mid X_3)$ might be low if the annotator is known to have high recall, since $L$ only includes 5 out of 20 nodes in $X_3$, as opposed to 4 out of 10 nodes in $X_2$ and 3 out of 5 nodes in $X_1$.

The second component of the score is $\mathbf{P}(X_i)$, which denotes the *goodness* of $X_i$ as a list. Let us evaluate the goodness on two measures, the *schema size* and the *repeating structure*. For $X_1$, since each cell in the first column corresponds to a name in a business listing, the cells between one name and next corresponds to an entire business listing. Thus, each business listing has 4 cells, so $X_1$ corresponds to a list with schema size 4. Also, the list has a perfect repeating structure, since each listing is one row. For $X_3$, each cell is a name, and thus, each business listing comprises of just one cell. Again, this has a perfect repeating structure, but the schema size is only 1. Given the domain knowledge that each business listing typically has name, address, phone, etc., the list $X_1$ is very unlikely. Thus, we expect $\mathbf{P}(X_3) < \mathbf{P}(X_1)$. List $X_2$, on the other hand, has a poor repeating structure, since the gap between first two elements ($n_1$ and $a_1$) is 0, while gap between next two elements ($a_1$ and $n_2$) is 2. So again, we expect $\mathbf{P}(X_3) < \mathbf{P}(X_2)$.

At the end, $w_1$ should be ranked as the top wrapper, since even though it misses two labels, it has good list properties. In Section 6, we describe our ranking model formally. But first, we address the issue of efficiently enumerating the candidate wrappers.

# 4. ENUMERATION

We have a set of labeled nodes $L$ and a wrapper inductor $\phi$. We use $\phi(L)$ to denote the wrapper learned from $L$. Since we work with a single type, $\phi(L)$ will also denote the set of nodes obtained by applying the wrapper on the given set of pages.

The *wrapper space* of a set of labels $L$, denoted $\mathcal{W}_\phi(L)$, is the set of unique wrappers defined by the subsets of $L$, i.e.,

$$\mathcal{W}_\phi(L) = \{\phi(L_1) \mid L_1 \subseteq L\}.$$

When $\phi$ is clear from context, we simply use $\mathcal{W}(L)$ to denote the wrapper space of $L$. In general, while the number of subsets of $L$ is exponential in size of $L$, $\mathcal{W}(L)$ is usually a small set, as we discussed in Section 3.

PROBLEM 1 (ENUMERATION). *Given a wrapper inductor $\phi$ and a set of labels L, output $\mathcal{W}(L)$ efficiently in time polynomial in the size of L and $\mathcal{W}(L)$.*

In general, for an arbitrary blackbox $\phi$, we cannot solve the enumeration problem without invoking $\phi$ on all possible subsets of $L$. However, any reasonable wrapper inductor algorithm has some basic properties that we can expect to hold and that we can exploit.

DEFINITION 1. *We say that a wrapper inductor $\phi$ is* well-behaved *if it satisfies the following properties:*

1. [FIDELITY] $L \subseteq \phi(L)$ *for all L,*
2. [CLOSURE] $\ell \in \phi(L) \Rightarrow \phi(L) = \phi(L \cup \{\ell\})$,
3. [MONOTONICITY] $L_1 \subseteq L_2 \Rightarrow \phi(L_1) \subseteq \phi(L_2)$.

The goal of a wrapper inductor is to generalize from a set of labeled examples. The fidelity property simply states that the generalization includes the original examples. E.g., TABLE from Example 1 trivially satisfies fidelity. The closure property states that the wrapper does not change if its output is given as additional training data. Monotonicity implies that if we add more labels, we do not extract fewer nodes. It is easy to see that TABLE is monotonic and closed, e.g., $\{n_1, n_2\}$ gets generalized to the entire first column that also includes $n_4$. If we start with $\{n_1, n_2, n_4\}$, then we will still get the entire first column. Thus, TABLE is well-behaved.

The properties in Definition 1 are very natural. As we show in Section 5, the Wien system [15] that generates wrapper rules in terms of prefixes and suffixes, and the XPath system [6] that generates xpath rules, are both well-behaved. In the next section, we give an efficient algorithm that takes a well-behaved $\phi$ as a blackbox, a set of labels $L$, and efficiently enumerates the wrapper space of $L$.

## 4.1 A bottom-up algorithm

We fix here a set of labels $L$. Given any subset $s \subseteq L$, define the *closure* of $s$, denoted $\breve{\phi}(s)$, as $\breve{\phi}(s) = \phi(s) \cap L$. The algorithm is very simple to describe and is called BottomUp. The pseudocode is given in Appendix D (Algorithm 1). It works bottom-up starting from the empty set. It maintains a list $\mathcal{Z}$ of subsets of $L$ that are candidates for learning wrappers. At each iteration, the smallest set in $\mathcal{Z}$ is chosen for expansion (step 4). It is expanded by one element in all possible ways (step 6), and for each expansion, the resulting wrapper is added to the set $\mathcal{W}$. Also, the closure of the expanded set is added back to $\mathcal{Z}$. Step 8 is the step that makes the algorithm efficient. If we do not take closures, then this would amount to a full enumeration over all $2^{|L|}$ subsets of $L$. We have included the run-time analysis of the algorithm below. First, we illustrate using an example.

EXAMPLE 2. *We revisit Example 1. The label set is given by $L = \{n_1, n_2, n_4, a_4, z_5\}$, and $\phi$ is the* TABLE *wrapper inductor that we have devised. Let $R_i$ denote row $i$, $C_i$ denote column $i$, and $T$ denote the whole table. When the algorithm starts, we have $\mathcal{W} = \emptyset$ and $\mathcal{Z} = \{\emptyset\}$. After $\emptyset$ is expanded, we get*

$$\mathcal{W} = \{\{n_1\}, \{n_2\}, \{n_4\}, \{a_4\}, \{z_5\}\},$$

$$\mathcal{Z} = \{\{n_1\}, \{n_2\}, \{n_4\}, \{a_4\}, \{z_5\}\},$$

*since each call to $\phi$ on a singleton label results in itself. Next we pick $\{n_1\}$ for expansion (we can pick any set since each of them is smallest). We get $\phi(\{n_1, n_2\}) = \phi(\{n_1, n_4\}) = C_1$, with corresponding $s_{\text{new}} = \{n_1, n_2, n_4\}$, and $\phi(\{n_1, a_4\}) = \phi(\{n_1, z_5\}) = T$, with the corresponding $s_{\text{new}} = L$. Thus, after expanding $n_1$, we get*

$$\mathcal{W} = \{\{n_1\}, \{n_2\}, \{n_4\}, \{a_4\}, \{z_5\}, C_1, T\},$$

$$\mathcal{Z} = \{\{n_2\}, \{n_4\}, \{a_4\}, \{z_5\}, \{n_1, n_2, n_4\}\}.$$

*When $\{n_2\}$ is expanded, it is removed from $\mathcal{Z}$ and it does not contribute any new set in either $\mathcal{W}$ or $\mathcal{Z}$. When $\{n_4\}$ is expanded, with $a_4$ it gives $\phi(\{n_4, a_4\}) = R_4$ with the corresponding $s_{\text{new}} = \{a_4, n_4\}$. After $\{n_4\}$ is done we get*

$$\mathcal{W} = \{\{n_1\}, \{n_2\}, \{n_4\}, \{a_4\}, \{z_5\}, C_1, R_4, T\},$$

$$\mathcal{Z} = \{\{a_4\}, \{z_5\}, \{n_4, a_4\}, \{n_1, n_2, n_4\}\}.$$

*No further set in $\mathcal{Z}$ contributes anything new, and successively gets removed from $\mathcal{Z}$. The final $\mathcal{W}$ contains the 8 wrappers as shown above. These are precisely the set of wrappers in Equation (2).*

We see that the algorithm correctly generates the wrapper space of $L$ in the above example. We prove that this is always the case.

THEOREM 1. BottomUp *is sound and complete, i.e., it precisely outputs the wrapper space of L.*

THEOREM 2. *Let k be the size of the wrapper space of L. Then,* BottomUp *makes at most $k \cdot |L|$ calls to the wrapper inductor.*

## 4.2 A top-down approach

While the bottom-up algorithm runs in time polynomial in the size of the wrapper space, it still, in the worst case, makes $k \cdot |L|$ calls to the wrapper, where $k$ is the number of unique wrappers. In this section, we look at non-blackbox algorithms for enumeration, where we use the form of the wrappers to devise optimal algorithms that generate the wrapper space with exactly $k$ calls to the wrapper. We define a special class of wrapper inductors, called *feature-based* inductors, as defined below.

A wrapper inductor $\phi$ is called *feature-based* if it can be expressed in the following form. Every node $n$ in a document is associated with a set of features $F(n)$, where a feature is an $(attribute, value)$ pair. Given a set $L$ of labeled nodes, $\phi(L)$ is defined as $\phi(L) = \{n \mid F(n) \supseteq \bigcap_{\ell \in L} F(\ell)\}$.

EXAMPLE 3. *Consider the* TABLE *wrapper from Example 1. We can view it as a feature-based wrapper in the following way. With each cell in the table, we associate two attributes,* `row` *and* `col`*, that denote the row number and column number of the cell. E.g., $n_1$ has features $\{(\texttt{row}, 1), (\texttt{col}, 1)\}$, $a_4$ has features $\{(\texttt{row}, 4), (\texttt{col}, 2)\}$ and so on. Thus, if $L = \{n_1, n_2, n_4\}$, $\phi(L)$ will take the intersection of their features, which is $\{(\texttt{col}, 1)\}$. Thus, it will generalize $L$ to the entire first column. Similarly, if $L = \{n_1, a_4\}$, the intersection of their features is empty. Hence, $\phi(L)$ is the whole table. One can check that with this feature set, the resulting $\phi$ is equivalent to* TABLE.

Though feature-based inductors seem to be a very special class of wrapper inductors, many current wrappers can be expressed in this form. For instance, as we show in Section 5, both the Wien system [15] that generates wrapper rules in terms of prefixes and suffixes, and the XPath system [6] that generates xpath rules, have a natural representation as a feature-based inductor.

Let $\mathcal{F}(L)$ be the union of all the features of all nodes in $L$. Let $\texttt{attrs}(L)$ denote all the set of unique attributes in $\mathcal{F}(L)$. Given $a \in \texttt{attrs}(L)$ and a subset $s$ of $L$, let $v_1, \ldots, v_t$ be the set of values that $a$ takes on $s$. Define $\texttt{subdivision}(s, a)$ to be the set $\{s_1, \ldots, s_t\}$, where $s_i$ contains all the nodes in $s$ that have a feature $(a, v_i)$. Note that a subdivision need not cover $s$, since there might be nodes in $s$ that do not have the attribute $a$ at all.

The enumeration algorithm is called TopDown. The pseudocode is given in Appendix D (Algorithm 2). It starts with the entire set of labels $L$, and starts creating smaller sets based on their features. We illustrate it on Example 1. Here $L = \{n_1, n_2, n_4, a_4, z_5\}$. There are two attributes, `row` and `col`. At the beginning, $\mathcal{Z} = \{L\}$. The `col` attribute divided it into following sets: $\{n_1, n_2, n_4\}$, $\{a_4\}$, and $\{z_5\}$. Thus, after processing `col`, we have $\mathcal{Z} = \{\{n_1, n_2, n_4\}, \{a_4\}, \{z_5\}, L\}$. When `row` is used, it divides $\{n_1, n_2, n_4\}$ into $\{n_1\}$, $\{n_2\}$, and $\{n_4\}$. It does not create any new sets in $\{a_4\}$ and $\{z_5\}$. For $L$, it creates $\{n_1\}$, $\{n_2\}$, $\{n_4, a_4\}$, and $\{z_5\}$. At the end, we have $\mathcal{Z} = \{\{n_1\}, \{n_2\}, \{n_4\}, \{a_4\}, \{z_5\}, \{n_4, a_4\}, \{n_1, n_2, n_4\}, L\}$. Recall that there are 8 unique wrappers for $L$ in Example 1 as given in Equation (2). When we call $\phi$ on each of the 8 subsets in $\mathcal{Z}$, we precisely get these 8 unique wrappers.

THEOREM 3. TopDown *makes exactly $k$ calls to the wrapper inductor, where $k$ is the size of the wrapper space.*

## 5. WRAPPER INDUCTORS

In this section, we analyze two wrapper induction algorithms from the literature in detail, in the context of our framework.

**The WIEN system.** The WIEN system from Kushmerick et al. [15, 14, 10] considers documents as a sequence of characters. It defines various class of wrapper languages, with the simplest being the LR wrappers. Given a set of labeled examples, the LR wrapper finds longest common strings preceding and following each of the examples. The wrapper consists of this pair of strings. Thus, the nodes obtained by the wrapper consists of all the minimal strings that are delimited by these pairs of strings. E.g., a wrapper consisting of the pair ("<td>", "</td>") will fetch all data items in all the tables in the documents. If extracting multiple types, there is a pair of delimiters for each type. There are various extensions of this basic languages, e.g., HLRT wrappers, which, in addition, have strings H and T that limit the context under which LR can be applied. Although we only present here the analysis for LR wrappers, the analysis also extends to HLRT and its other variants.

THEOREM 4. LR *is a well-behaved wrapper inductor.*

The above result is not surprising. However, it comes as a surprise that LR *can* be expressed as a feature-based wrapper. To see this, consider the following set of features for each substring $\ell$ in the document: for each $k$ there is an attribute $L_k$ with value equal to the string of length $k$ immediately preceding $\ell$, and an attribute $R_k$ with value equal to the string of length $k$ immediately following $\ell$. It is easy to see that the resulting feature-based wrapper is equivalent to LR. E.g., given a set of labeled strings, if their longest preceding string has length $k$, then they will share common features corresponding to $L_1, \ldots, L_k$, and so on.

Since we can express LR as a feature-based wrapper, we can apply the TopDown algorithm to enumerate its wrapper space. However, note that in this feature space, each label will have a large number of features, equal to the length of the document. The charm of the algorithm is that we do not need to construct the feature space, as long as we can efficiently implement the `subdivision` routine. For LR wrappers, it is easy to implement `subdivision`, and we omit the details here for lack of space.

**The XPATH wrapper.** Several papers consider the approach of looking at documents as trees given by their parsed HTML mark-up, and using xpaths to extract information. The algorithm we describe here is from Dalvi et al. [6], which we refer to in this paper as XPATH [2]. XPATH considers a simple fragment of the xpath language, consisting of child edges (/), descendant edges (//), attribute filters (`[@fontsize=2]`) and child number filters (`td[2]`). For example, the following xpath

$$//\texttt{div}[@\texttt{class} =' \texttt{content}']/\texttt{table}[1]/\texttt{tr}/\texttt{td}[2]/\texttt{text}(), \quad (3)$$

extracts the text from the second column of each row in the first table from a specific `div`. The wrapper induction consists of starting from the "`//*`" xpath and specializing it till the precision cannot be further improved, while maintaining a recall of 1. One can verify:

THEOREM 5. XPATH *is a well-behaved wrapper inductor.*

Again, while not intuitive, XPATH can in fact be expressed as a feature-based inductor as follows. For each node $n$ in the document, we consider the path from $n$ to the root, and look at the properties at each position in the path. We will omit the details and the formal construction owing to lack of space, but instead illustrate the set of features for the xpath in Equation (3). It has `td[2]` at position 1, `tr` at position 2, `table[1]` at position 3 and so on. Its set of features is given by

{(1:tagname, `td`), (1:childnumber, 2), (2:tagname, `tr`), (3:tagname, tagname), (3:childnumber, 1), ...}.

Given a set of labeled nodes in this representation, one can construct an xpath by taking the intersection of the features of all the nodes. The formal details are omitted. The above representation enables us to express XPATH as a feature-based wrapper using a small number of features, and TopDown algorithm can be used to enumerate the wrapper space for XPATH.

Note that extractors based on machine-learning methods need not be well-behaved since they may not satisfy monotonicity. Furthermore, if a wrapper inductor already handles noise perhaps using different techniques, it may not be well-behaved.

## 6. RANKING MODEL

In this section, we describe the ranking model we use to assign a probability to each wrapper in the set of enumerated wrappers. We assume that we are given a wrapper $w$. Let $X$ be the set of nodes returned by $w$ on the given set of webpages. The score of $w$, according to Equation (1), is given by $\mathbf{P}(L \mid X)\mathbf{P}(X)$, where the first term models the annotation process and the second models the web publication process. Note that the actual language used to express $w$ does not matter, as the score of a wrapper only depends on its output. We describe each component of the score below.

**Annotation process.** $\mathbf{P}(L \mid X)$ is the probability that the annotator outputs the labeled set $L$ given that the correct list is $X$. We assume a simple annotation model, where the annotator looks at each node in the document, and independently decides whether to include it in $L$. E.g., a dictionary-based annotator might label a node if it belongs to the dictionary, and a zipcode-based annotator might label a node if it consists of a sequence of 5 digits.

We characterize an annotator by two parameters, $p$ and $r$. For each node in the correct list $X$, it is added to the label set $L$ with probability $r$. Also, for each node not in $X$, it is also added to $L$ with probability $1 - p$. An annotator with $p = r = 1$ corresponds to the perfect annotator that labels all the correct nodes and only correct nodes.

[2]While XPATH [6] focuses on robustness of wrappers w.r.t changes in website, this aspect is orthogonal to our setting, and we are only concerned here with their wrapper induction algorithm as described in Section 5 in their paper.

We can compute $\mathbf{P}(L \mid X)$ as follows. Let $A$ be the set of all nodes other than $X$ in the given set of webpages. Let $X_1 = X \cap L$, $X_2 = X \setminus L$, $A_1 = A \cap L$, $A_2 = A \setminus L$. Then,

$$\begin{aligned}\mathbf{P}(L|X) &= r^{|X_1|} \cdot (1-r)^{|X_2|} \cdot (1-p)^{|A_1|} \cdot p^{|A_2|} \\ &= \left(\tfrac{r}{1-p}\right)^{|X_1|} \cdot \left(\tfrac{1-r}{p}\right)^{|X_2|} \cdot (1-p)^{|A_1 \cup X_1|} \cdot p^{|A_2 \cup X_2|}.\end{aligned}$$

We observe that while $X$ and $A$ differ for each wrapper, $A_1 \cup X_1$, which is equal to $L$, remains invariant. Similarly, $A_2 \cup X_2$, which is equal to the complement of $L$, again remains invariant over different wrappers. Thus, we have

$$\mathbf{P}(L|X) \propto \left(\tfrac{r}{1-p}\right)^{|L \cap X|} \cdot \left(\tfrac{1-r}{p}\right)^{|X \setminus L|}. \tag{4}$$

We use Equation (4) to model the annotation process. The reader can notice that the term $p$, as we have defined, is *not* the precision of the annotator, though the two are closely related. Assuming that $1 - p < r$, Equation (4) is indeed maximized when $X = L$. If $1 - p > r$, it means that the annotator picks wrong nodes with higher probability than right nodes, and indeed Equation (4) is maximized when $X$ is complement of $L$. In such a case, equivalently, we can flip the output of the annotator and use it instead.

**Web publication model.** The second term, $\mathbf{P}(X)$, models the "goodness" of the list in terms of its structure. There are several work that model the web publication process [2, 5], and try to infer the grammar that generated the given set of webpages. We can measure the goodness of a list by studying its grammar. A regular, structurally repeating list results in a simple grammar, while an irregular list requires a complex grammar.

While we can adapt any of these grammar inference technique to our setting, there are differences unique to our setting that we can exploit to device simpler and more efficient alternate techniques. First, we do not use the grammar directly for extraction, but only use it as a means to rank wrappers. Thus, we only need the characteristics of the grammar and the actual grammar is not important for us. Second, the grammar inference techniques work on unsupervised pages, and try to segment the pages into structurally similar segments that align well. On the other hand, since we are evaluating the goodness of a list, we already have the target list $X$ that we can use for segmentation. Thus, we only need to check the structural similarity of the segments. We present below our technique that exploits these two properties.

We consider the parsed DOM trees of the websites, replacing each piece of text with a special node called `<#text>`, since we are only concerned with the structure and not the content. Next we use the nodes in $X$ as record boundaries to obtain record segments. Note that this might lead to shifted record segments, since elements in $X$ might occur in the middle of the records. E.g., if we have a sequence

$$a_1 n_1 z_1 p_1 \; a_2 n_2 z_2 p_2 \; a_3 n_2 z_3 p_3 \ldots$$

and if $X$ contains all the $n_i$, then the record segments will be

$$(n_1 z_1 p_1 a_2), (n_2, z_2, p_2, z_3)$$

and so on, which are cyclically shifted, but the structural similarity of the records is preserved, which is what matters to us. This trick allows us to obtain record segments easily. In the parsed DOM trees, this segmentation is performed by doing an pre-order traversal of tree nodes from one consecutive element of $X$ to the next. See Figure 7 in the Appendix for an illustration of this process on a simple HTML page. The result is a list of record segments, where each segment is a fragment of HTML tree.

## 6.1 Features

To compute $\mathbf{P}(X)$, we define a set $F$ of features on the list of record segments, along with a probability distribution on the values for each feature, and compute $\mathbf{P}(X)$ as $\prod_{f \in F} \mathbf{P}(f)$. Several useful features can be defined that encapsulate the domain knowledge. In this paper, we only consider the following two features:

(1) *Schema size*: This is the number of text nodes in the longest common substring between pairs of segments; notice that this is an approximation for the number of text attributes that are present in every record in the list. E.g., as discussed in Section 3, if we are extracting addresses, then it is unlike to see a schema size 1 or a schema size of 50. We learn the schema size distribution for each domain from a sample consisting of websites in the domain.

(2) *Alignment*: This is the maximum pairwise edit distance between pairs of segments; this feature is meant to capture how well the records align. Once again, we learn the alignment distribution for each domain from a sample of websites in the domain. An alignment score of 0 corresponds to a perfect list. Section 3 gives examples of wrappers leading well-aligned and badly-aligned lists.

Note that these features are applicable in all domains, though their value distribution might be domain-specific. It is possible to use features specific to a domain, e.g. every address has a zipcode and a business typically has 1 or 2 phone numbers. In our experiments, we find that even the two domain-independent simple features described above are sufficient to identify good lists.

To learn the probability distribution on the values of features, we take a small sample of websites, look at the list of segments on each website and learn the distribution. Since both schema size and alignment are discrete valued features, we use the kernel density methods that learn a smooth distribution from finite data samples.

# 7. EXPERIMENTS

We first describe the setup of our experiments, our test data, and the wrapper induction techniques we use for evaluation.

**Datasets.** We created two datasets in two different domains. In the first dataset, the task is to extract store names from dealer locator pages of various businesses. We compiled a list of 330 businesses over various categories like furniture, home appliances, and electronics. Each business had a dealer locator webpage, where, by automatically filling the US zipcode, we obtained the list of dealers for each business for each zipcode. Note that pages for different zipcodes from the same business have the same structure, while different businesses have completely different websites. We want to automatically learn wrappers for each of the 330 websites. We call this dataset DEALERS.

In the second dataset, the task is to extract track names from music albums. We crawled 15 different discography sites, where each site contained structurally similar pages for albums along with their track information listing. Again, we want to automatically learn wrappers for each of the website. Figure 8 in Appendix contains the details of this dataset, which can be used for repeatability of the experiments. We call this dataset DISC.

**Annotators.** Our annotator for DEALERS is based on the Yahoo! Local database of about 600,000 business names, e.g., `Home Depot` and `BestBuy`, which overlaps with some of the business names for each directory. We annotate a text node if it contains an exact mention of a business name from our database. The resulting annotator has a 0.95 precision and 0.24 recall. The errors stem from a business names matching street addresses and product descriptions. For the DISC dataset, we compiled a list of 11 popular albums along with their track information, spanning various genres and decades. We expect any discography website to have at least a few of these albums, if not all, allowing us to automatically label and learn a wrapper for the website. Again, we look for exact track names on the webpages. The annotator has precision 0.8 and recall 0.9. Note that the recall is only measured w.r.t to pages with at least one annotation. Thus, a recall of 0.9 does not mean that we had 90% of the albums in our database (we only had 11). It means that on the pages corresponding to the albums from our database, we locate around 90% of the track titles. Errors in annotations stem from track titles matching album titles, or present inside album descriptions/user comments on the page.

For evaluation of our methods, we manually created the correct extraction rule for each website in DEALERS and DISC.

**Algorithms.** We experimented with two different wrapper induction algorithms, XPATH and LR, as defined in Section 5. We have implemented our framework in Java. All our experiments are run on an Intel 2.13GHz machine running Linux with 4GB RAM. We use the `tidy` (`http://sourceforge.net/projects/jtidy`) utility to clean up and parse HTML pages.

**Learning the model parameters.** For each domain, the probability distribution of the two features, namely, schema size and alignment, and the $p$ and $r$ of the annotators are learned from a sample of half the websites.

## 7.1 Wrapper enumeration

In this experiment, we evaluate the performance of TopDown and BottomUp wrapper enumeration algorithms, and compare them with the naive algorithm that does an exhaustive search over all possible subsets of $L$.

Figure 2(a) shows the number of wrapper calls made by the three approaches for the LR wrappers. All the websites are arranged along the $x$-axis in increasing order of the TopDown time, and for each website there are three points corresponding to the three approaches. The naive method is not plotted when it gets too large. Figure 2(b) shows the same graph for the XPATH wrappers. We see that both TopDown and BottomUp show excellent efficiency as compared to the naive enumeration, with BottomUp making an order of magnitude more calls to the wrapper.

Figure 2(c) shows the physical running times of the two algorithms for XPATH wrapper. The naive algorithm was not run as it was prohibitively expensive. Again, we see a similar trend. The TopDown algorithm was really efficient and finished in under a second for most of the websites. The BottomUp was again an order of magnitude slower, but still finished in under a couple of minutes for most websites.

## 7.2 Ranking

In this experiment we evaluate the improvement in extraction quality obtained using our techniques. For each wrapper inductor, we compare two algorithms: (1) the naive algorithm (NAIVE) that directly runs the inductor on the given set of noisy annotations, and (2) our noise-tolerant wrapper framework (NTW).

Note that we do not need to evaluate the TopDown and BottomUp algorithms separately for ranking, since both of them simply enumerate the wrapper space, which is orthogonal to performance of the ranking algorithm.

Figure 2(d) shows the precision, recall, and the $f_1$-measure, which is the harmonic mean of the precision and recall, for the naive algorithm and our methods using XPATH wrappers. We observe a nearly perfect precision and recall using our noise-tolerant wrapper framework. The naive algorithm has a perfect recall. This is because, due to noise in the training, it over-generalizes the matching rule, leading to a very low precision. We are able to improve the precision to 1 with a negligible drop in the recall.

Figure 2(e) shows the same graph on LR wrappers. It shows the same trend, but the effect is more pronounced. Since LR wrappers are less expressive than XPATH wrappers, the over-generalization due to the noise is more severe, leading to a really low precision of the naive method. Secondly, we see that the accuracy of the noise-tolerant framework, while quite high, is still only 90% while XPATH performs close to 100% on the same dataset. This is again due to the fact that LR wrappers are not as expressive as XPATH wrappers and for some of the websites, a perfect LR wrapper does not exist. Thus, LR wrappers cannot achieve perfect accuracy on this dataset even when there is zero noise.

Figures 2(f) and 2(g) shows the same graphs on the DISC dataset. We see that our noise tolerant framework achieves a perfect precision and recall on this dataset for both wrapper inductors.

## 7.3 Ranking model components

Our ranking model has two components, the labeling error term $\mathbf{P}(X \mid L)$, and the term $P(X)$ that models the goodness of the list. In this experiment, we analyze the contribution of each component to the accuracy of the wrapper induction. We define two variants of NTW corresponding to the two components, denoted NTW-L and NTW-X respectively, where each variant only takes into account one component of the ranking. Figures 2(h) and 2(i) show the accuracy of each variant. We see that none of the individual components can account for the total accuracy of the technique. Furthermore, we observe that the individual contribution of the components differ significantly for XPATH and LR wrappers. For XPATH wrappers, simply using the labeling errors almost takes us all the way to the maximum accuracy, while for LR, labeling errors by themselves do not help much.

## 7.4 Effect of annotators

In this experiment, we study the ability of our techniques to handle annotators with different precision and recall characteristics. So far, we have looked at two annotators : the DEALERS annotator with a 0.95 precision and a 0.24 recall, and the DISC annotator with a 0.81 precision and a 0.9 recall.

Here, we vary the precision/recall of annotators in a controlled way, and study the effect on our techniques. For this study, we consider the following annotator: it takes the set of corrects nodes as input. For each correct node, it annotates it with probability $p_1$. Also, for each incorrect node, it annotates it with probability $p_2$. Note that the expected recall of the annotator is precisely $p_1$. The expected precision of the annotator, assuming $n_1$ correct nodes and $n_2$ incorrect nodes, is $n_1p_1/(n_1p_1 + n_2p_2)$. Thus, by controlling $p_1$ and $p_2$, we can construct annotators with any desired precision and recall. For each website, we use this annotator to annotate 25 webpages.

We take the DEALERS dataset and the XPATH wrappers, and study the accuracy ($F_1$ measure) of our techniques as a function of precision/recall of annotators. The result is shown in Table 1. The region with more than 90% accuracy is highlighted. We observe that our techniques have a very effective noise tolerance over a broad range of annotators, and have very modest precision and recall requirements.

# 8. CONCLUSIONS

We present a generic framework to learn wrappers from noisy training data. This framework combines supervised wrapper induction techniques with domain knowledge and unsupervised grammar induction techniques and allows to (i) plug in any wrapper language and wrapper induction system and (ii) learn wrappers effectively from automatically and cheaply obtained noisy training data.

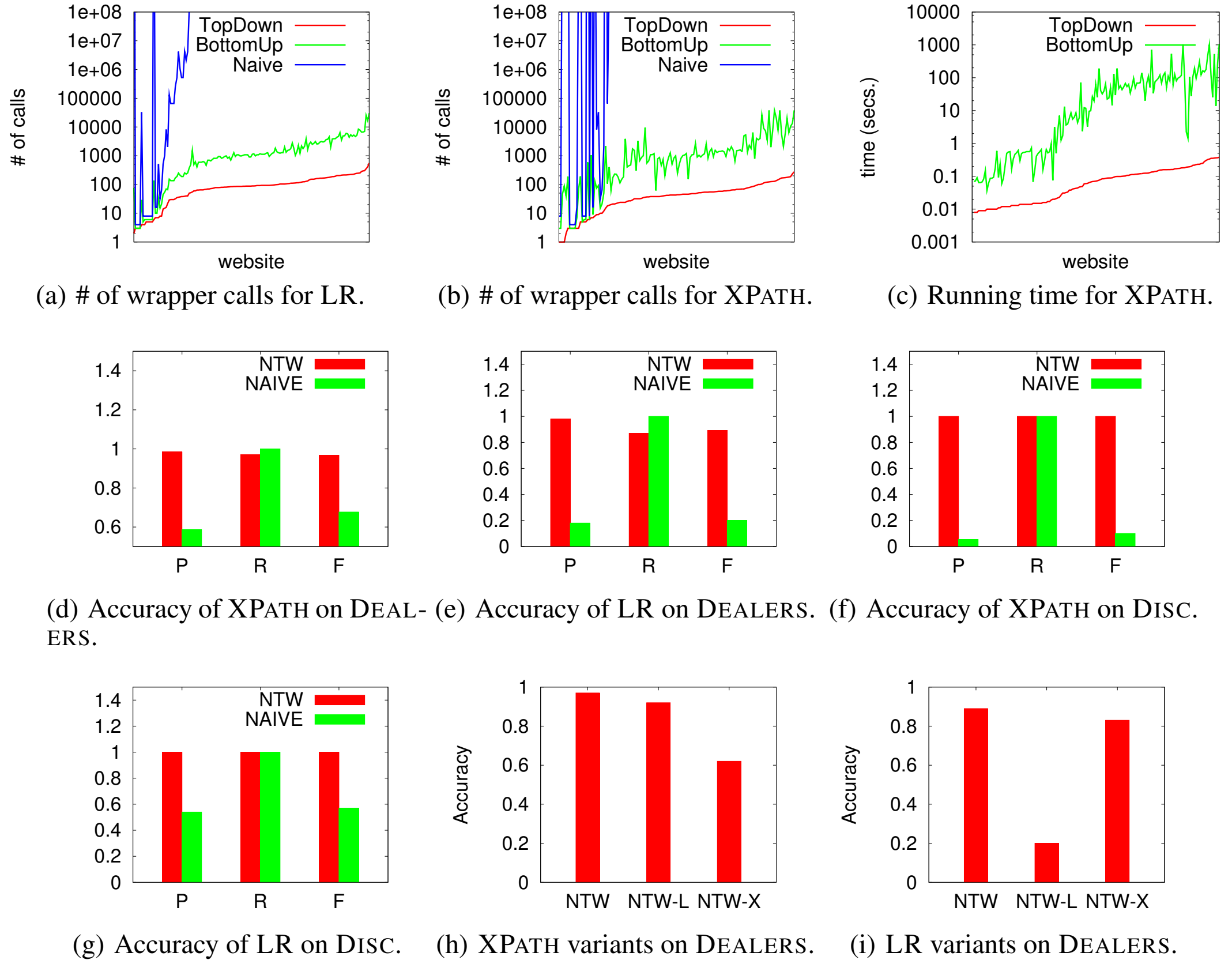

(a) # of wrapper calls for LR. (b) # of wrapper calls for XPATH. (c) Running time for XPATH.

(d) Accuracy of XPATH on DEALERS. (e) Accuracy of LR on DEALERS. (f) Accuracy of XPATH on DISC.

(g) Accuracy of LR on DISC. (h) XPATH variants on DEALERS. (i) LR variants on DEALERS.

**Figure 2: Evaluation.**

| p \ r | 0.05 | 0.1 | 0.15 | 0.2 | 0.25 | 0.3 |
|---|---|---|---|---|---|---|
| 0.1 | 0.41 | 0.67 | 0.72 | 0.75 | 0.73 | 0.73 |
| 0.3 | 0.56 | 0.82 | 0.88 | 0.89 | **0.93** | **0.93** |
| 0.5 | 0.67 | 0.82 | 0.88 | **0.92** | **0.93** | **0.95** |
| 0.7 | 0.69 | 0.85 | **0.92** | **0.93** | **0.95** | **0.95** |
| 0.9 | 0.73 | 0.88 | **0.93** | **0.94** | **0.96** | **0.97** |

**Table 1: Accuracy of NTW as a function of annotator**

# APPENDIX

# A. MULTI-TYPE EXTRACTION

In the main paper, for ease of exposition, we only presented techniques for *single-type extraction*. However, our techniques are equally applicable for the general *multi-type extraction* problem, and we present its treatment here.

## A.1 Methodology

A multi-type wrapper inductor seeks to extract a list of records, where a record consists of a set of types. An example is the task of extracting business listings consisting of `name`, `address`, and `phone` from dealer locator pages. Each type has a corresponding annotator for it. Learning a multi-type wrapper is a significantly harder task than learning wrappers for individual types, since in addition, the wrapper needs to *assemble* records by putting together corresponding data items. The task of assembling records is further complicated when some of the fields are missing from certain records, e.g., a business name with no phone number.

While wrapper induction for multi-type extraction from clean labeled data is a hard problem and a topic of continuing research, our focus here is to make such an extractor noise-tolerant. As we show below, we can extend our techniques is a straightforward way.

**Enumeration.** The algorithms described in Section 4 can also be used for enumerating multi-type wrappers. The labeled nodes now additionally have a type associated with them. The type information is passed along when calls are made to the wrapper inductor.

**Ranking.** There are two components in ranking, $\mathbf{P}(L \mid X)$ and $\mathbf{P}(X)$. We assume that we have multiple annotators, one for each type, each with a *precision* and *recall* associated with it, as defined in Section 6. The $\mathbf{P}(L \mid X)$ term, which models the annotation errors, is computed by multiplying the annotation errors of each individual annotator, computed independently using Eq. (4). For the $\mathbf{P}(X)$ term, we obtain the DOM tree segments corresponding to individual records by using any of the types as a record boundary. Then we compute the schema size and the edit distance between segments; for the latter, we enforce the additional constraint that nodes corresponding to each type align with each other.

## A.2 Experiments

We use the DEALERS dataset described in Section 7 that consists of dealer locator pages of various businesses. To evaluate the effectiveness of our techniques on multi-type wrapper inductors, we developed an inductor operating on two types: `name` and `zipcode`. The inductor takes a set of nodes labeled as names and a set of nodes labeled as zipcodes, learns xpath rules for each type and constructs records by assembling together the interleaved names and zipcodes. The wrapper produces empty results on a page if it cannot assemble records successfully.

We also created two annotators: the *name annotator*, which is based on a database of business names as described in Section 7, and the *zipcode annotator* which is a regular expression identifying five-digit US zipcodes. In addition to the noise in names that we observed previously, there is noise in the zipcode labels, where match all five-digit street address, as well as text from page headers/footers.

When given clean manual labeled examples, the wrapper inductor was able to learn correct rules for names and zipcodes and successfully construct records. However, the effect of noise was more dramatic than the single-type extraction case, since an error in learning even one of the types made the wrapper fail in successfully assembling records. Thus, automatically learning a multi-type extractor warranted an even stronger need for noise-tolerance.

| | |
|---|---|
| 1. | `bizrate.com` |
| 2. | `shopping.yahoo.com` |
| 3. | `pricegrabber.com` |
| 4. | `google.com/products` |
| 5. | `shopper.cnet.com` |
| 6. | `puremobile.com` |
| 7. | `letstalk.com` |
| 8. | `mysimon.com` |
| 9. | `tigerdirect.com` |
| 10. | `shopping.com` |

**Figure 4: List of websites used in PRODUCTS dataset.**

Figure 3(a) shows the precision, recall, and the $f_1$-measure for the naive algorithm (NAIVE), which involves running the wrapper inductor directly on the noisy labeled data, and our noise-tolerant wrapper (NTW). The figure shows the dramatic effect of noise on the naive multi-type wrapper. The recall (and hence the $f_1$-measure) is close to 0, since NAIVE learns imperfect wrapper for either name or zipcode (or both) and fails to assemble records. On the other hand, NTW is able to achieve precision and recall close to 1.

We also observed that while NAIVE for multi-type extraction behaves significantly worse than in single-type extraction case, NTW actually performs slightly better. Figure 3(b) shows the average accuracy ($f_1$-measure) for names and zipcodes given by the noise-tolerant multi-type wrapper, as compared to the accuracy of each type when extracted singly. We see that when extracted jointly, the accuracy of zipcodes matches the single extraction, while the accuracy of names is slightly higher. The reason for NTW performing as good (or slightly better) in multi-type setting is that annotations of each type, even though not clean, help in evaluating the wrapper for the other type, since their joint alignment is measured during ranking. In NAIVE, in stark contrast, the imperfect annotations of one type adversely affect the other type since records cannot be assembled correctly.

# B. OTHER EXPERIMENTS

## B.1 Products domain

We created a third dataset, called PRODUCTS, consisting of shopping websites. We focused on websites selling cellphones, and the task is to extract all cellphones sold on each website. We crawled 10 different websites, listed in Figure 4. We constructed the dictionary by looking at five popular brands, and consulting the Wikipedia pages that list all the cellphone models for each of these brands. The total size of the dictionary was 463. Figure 3(c) shows the precision, recall, and the $f_1$-measure of NAIVE and NTW methods using XPath wrappers on this dataset. We note that the graphs show a behavior similar to DEALERS and DISC datasets.

## B.2 Single-entity extraction

Our techniques are also applicable for the easier problem of *single-entity extraction*. In single-entity extraction problem, each page contains only one entity of interest, as opposed to a list of entities. Many of the classic wrapper induction systems only handle single-entity extraction.

There are some salient differences between single-entity extraction and list extraction. First, in the absence of lists, we cannot exploit the repeated structure within a page to identify good lists. The $\mathbf{P}(X)$ component of our ranking, that models the goodness of a list, is not applicable for single-entity extraction. On the other hand, the constraint of having a single entity on a page greatly simplifies the problem. In list extraction, there is a *tension* between two conflicting objectives : (1) expand the set of annotations to the entire list in the pages, and (2) contract the set of annotations to account for noise. This tension is not present when extracting single entities, since we are not expanding the set of annotations in a page.

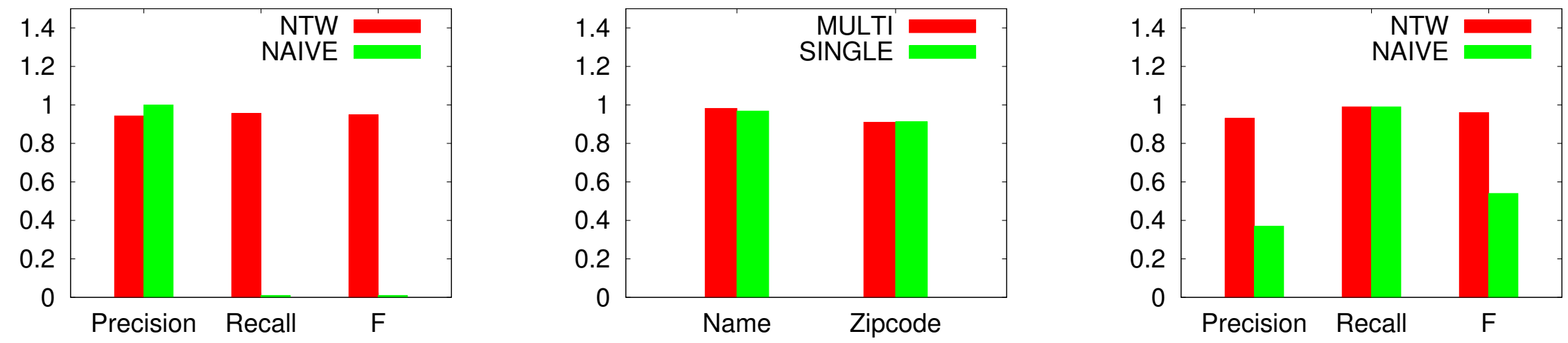

(a) Accuracy of multi-type extractor. (b) Multi-type vs. single-type extraction. (c) Accuracy of XPath on PRODUCTS.

**Figure 3: More evaluation.**

We can define the single-entity extraction problem as : find a wrapper that extracts a single entity from each webpage and covers most number of annotations. Note that covering the most number of annotations is equivalent to maximizing $\mathbf{P}(L \mid X)$.

Our techniques provide a direct way to solve the single-entity extraction problem (with noisy annotations) for any wrapper algorithm. We simply enumerate the set of distinct wrappers, using any of the applicable method, discard the wrappers that extract multiple items from a single page, and then pick the wrapper that covers the largest number of labels. The intuition is that any wrapper trained on a subset of annotations containing errors will most likely overgeneralize, match multiple nodes from a single page, and will be discarded.

We evaluated the performance of single-entity extraction on the DISC dataset. The task is to extract the album title from each page on each website. We use the same set of albums as our seed database for annotations. The annotator is very noisy, since album names are present at several locations in a page, including title tracks in some cases, as well as comments and reviews. However, the noise-tolerant wrapper was able to learn the correct wrapper in all the websites. In some of the websites, the system returned multiple wrappers with the same rank at the top, i.e. multiple wrappers matching one label on each page. These websites had multiple correct wrappers: there are multiple (consistent) locations containing album names in each page, like in the `head` tag, the `meta` tag, as well as album details tab.

In general, we expect single-entity extraction to be very amenable to noise in our framework.

# C. PROOFS

THEOREM 6. BottomUp *is sound and complete, i.e., it precisely outputs the wrapper space of L.*

PROOF. The soundness of the algorithm is obvious, since each wrapper in $\mathcal{W}$ is produced by a call to the wrapper inductor on some subset of $L$.

To show completeness, suppose on the contrary that a wrapper $w = \phi(s_0)$ is never produced. Let $s_1$ be the largest subset of $\breve{\phi}(s_0)$ that was ever added to $\mathcal{Z}$ in step 11, and let $s_1 = \breve{\phi}(s \cup \ell)$, where $s$ and $\ell$ are from step 8.

Suppose $s_1 = s_0$. Since $s_1 = \breve{\phi}(s \cup \ell)$, by the closure property of $\phi$, we have $\phi(s_1) = \phi(s \cup \ell)$. But $\phi(s \cup \ell)$ was added to $\mathcal{W}$ in step 9. Thus, if $s_1 = s_2$, this implies that $\mathcal{W}$ contains $\phi(s_0)$ which is a contradiction.

Hence, we can assume $s_1 \subset s_0$. Let $\ell_1$ be any label in $s_0 \setminus s_1$. Since $s_1$ was in $\mathcal{Z}$, at some point it was expanded, and $\ell_1$ was added to $s_1$. So consider $s_{\text{new}} = \breve{\phi}(s_1 \cup \ell_1)$.

Since $s_1 \cup \ell_1 \subseteq s_0$, by the monotonicity property, $\phi(s_1 \cup \ell_1) \subseteq \phi(s_0)$, which implies $(s_1 \cup \ell_1) \subseteq \breve{\phi}(s_0)$. Thus, we have a set $s_{\text{new}}$, such that $s_{\text{new}} \subseteq \breve{\phi}(s_0)$, and which is strictly larger than $s_1$ since it contains $\ell_1$ by the fidelity property. This contradicts the definition of $s_1$, and proves the completeness of the algorithm. □

Before we present the run time analysis of BottomUp, we need some notation. Call a subset $s$ of $L$ *closed* if $\breve{\phi}(s) = s$.

LEMMA C.1. *For any s, $\phi(s) = \phi(\breve{\phi}(s))$. Also, for any s, $\breve{\phi}(s)$ is closed.*

The lemma follows easily from the definition of $\breve{\phi}(s)$ and the closure property of $\phi$.

LEMMA C.2. *The size of the wrapper space of L is equal to the number of closed subsets of L.*

PROOF. Consider the set of wrappers $S = \{\phi(s) \mid s \text{ is closed}\}$. We will show that $S$ includes all the wrappers in the wrapper space of $L$. For given any wrapper $w = \phi(s_o)$, where $s_o$ is any set, $\phi(s_0) = \phi(\breve{\phi}(s_0))$ because of Lemma C.1. Also, $\breve{\phi}(s_0)$ is a closed set, again by Lemma C.1. Thus, $S$ contains $\phi(\breve{\phi}(s_0)) = w$.

Further, each closed set contributes a unique wrapper to $S$. This is because if $s_1$ and $s_2$ are closed sets and $s_1 \neq s_2$, this implies $\breve{\phi}(s_1) = s_1 \neq s_2 = \breve{\phi}(s_2) \Rightarrow \phi(s_1) \neq \phi(s_2)$. □

THEOREM 7. *Let k be the size of the wrapper space of L. Then,* BottomUp *makes at most $k \cdot |L|$ calls to the wrapper inductor.*

PROOF. First, we observe that the only sets we add to $\mathcal{Z}$ are of the form $\breve{\phi}(s \cup \ell)$, which, by Lemma C.1, are closed. Thus, $\mathcal{Z}$ only contains closed sets. Further, once a set is removed from $\mathcal{Z}$, it is never added back. This is because we always pick the smallest set for expansion in step 4. Thus, if a set $s$ is removed from $\mathcal{Z}$ at some step, all the remaining sets in $\mathcal{Z}$ are at least as big as $s$. Thus, any set ever added back to $\mathcal{Z}$ after this step is strictly bigger in size than $s$ (since at least one new label is added during step 8). So $s$ can never be added back to $\mathcal{Z}$ at a later stage.

Consider the set $s \cup \ell$ on which $\phi$ is called in step 7. Since $s$ comes from $\mathcal{Z}$, it is a closed set. Hence, $\phi$ is called at most once for each $(s, \ell)$ pair, where $s$ is a closed set and $\ell$ is a label. Since number of closed sets is $k$, BottomUp makes at most $k \cdot |L|$ calls to the wrapper $\phi$. □

# D. PSEUDOCODE

The pseudocode for BottomUp is shown in Algorithm 1 and for TopDown is shown in Algorithm 2.

**Algorithm 1** BottomUp

**Require:** $L$: a set of training labels, and $\phi$: a well-behaved black-box wrapper inductor.
**Ensure:** $\mathcal{W}$: the wrapper space of $L$.

$\mathcal{W} \leftarrow \emptyset$ {initialize wrapper space}
$\mathcal{Z} \leftarrow \{\emptyset\}$ {initialize set of label subsets}
**while** $\mathcal{Z} \neq \emptyset$ **do**
  $s \leftarrow$ smallest set in $\mathcal{Z}$
  $\mathcal{Z} \leftarrow \mathcal{Z} \setminus \{s\}$
  **for** (each label $\ell \in L \setminus \{s\}$) **do**
    $w = \phi(s \cup \ell)$
    $s_{\text{new}} = \breve{\phi}(s \cup \ell)$
    $\mathcal{W} \leftarrow \mathcal{W} \cup \{w\}$
    **if** ($s_{\text{new}} \neq L$) **then**
      $\mathcal{Z} \leftarrow \mathcal{Z} \cup \{s_{\text{new}}\}$
    **end if**
  **end for**
**end while**
**return** $\mathcal{W}$

**Algorithm 2** TopDown

**Require:** $L$: a set of training labels
**Ensure:** $\mathcal{W}$: the wrapper space of $L$.

$\mathcal{W} \leftarrow \emptyset$
$\mathcal{Z} \leftarrow \{L\}$
**for** (each attribute $a \in \texttt{attrs}(L)$) **do**
  **for** (each set $s \in \mathcal{Z}$) **do**
    $\{s_1, \ldots, s_t\} = \texttt{subdivision}(s, a)$
    $\mathcal{Z} \leftarrow \mathcal{Z} \cup \{s_1, \ldots, s_t\}$
  **end for**
**end for**
**for** (each set $s \in \mathcal{Z}$) **do**
  $\mathcal{W} \leftarrow \mathcal{W} \cup \{\phi(s)\}$
**end for**
**return** $\mathcal{W}$

# E. SUPPLEMENTARY FIGURES

Figure 5 and Figure 6 contain sample business listings pages. Figure 7 illustrates finding boundaries of records spanned by a given wrapper. Figures 8 and 9 contain the list of websites and albums used in DISC dataset.

# F. RELATED WORK

There are two existing lines of work related to our problem, which we describe here and outline their shortcomings. The first line of work is the non-wrapper based information extraction techniques that do not use the structure of the webpages, e.g., techniques from natural language processing [8], machine-learned probabilistic models like Conditional Random Fields (CRFs). While the techniques are not domain-centric, they are still *source-centric*, i.e., they perform poorly if the test data differs in *characteristic* from the training data. It is not possible to train, or even represent, a probabilistic model that can capture the full diversity of the entire web.

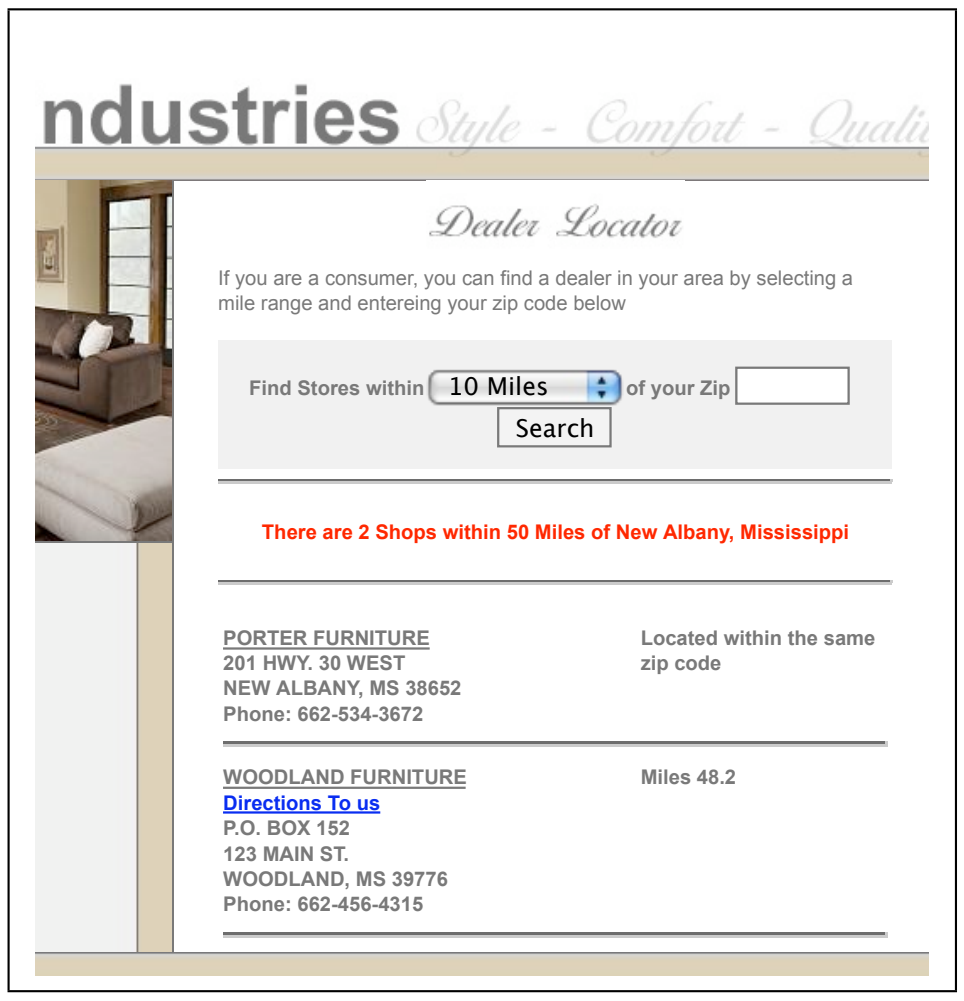

**Figure 5: Business listings webpage.**

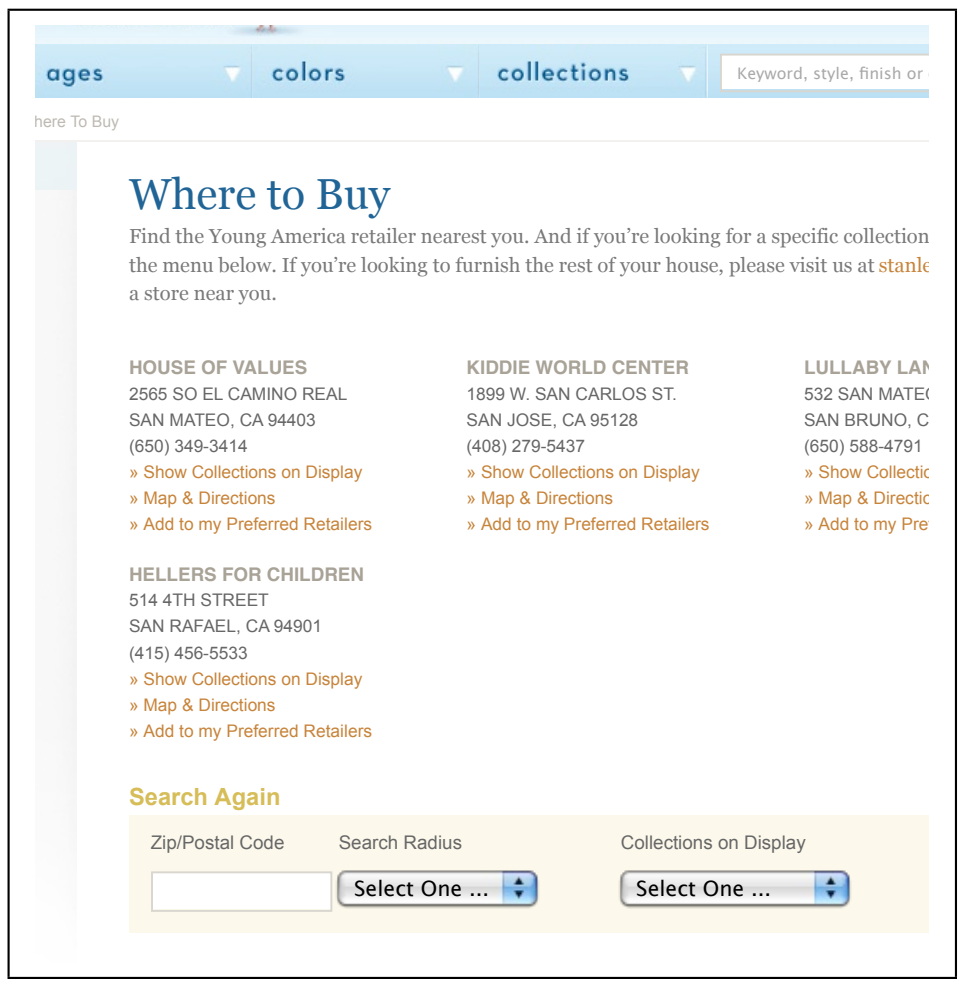

**Figure 6: Business listings webpage.**

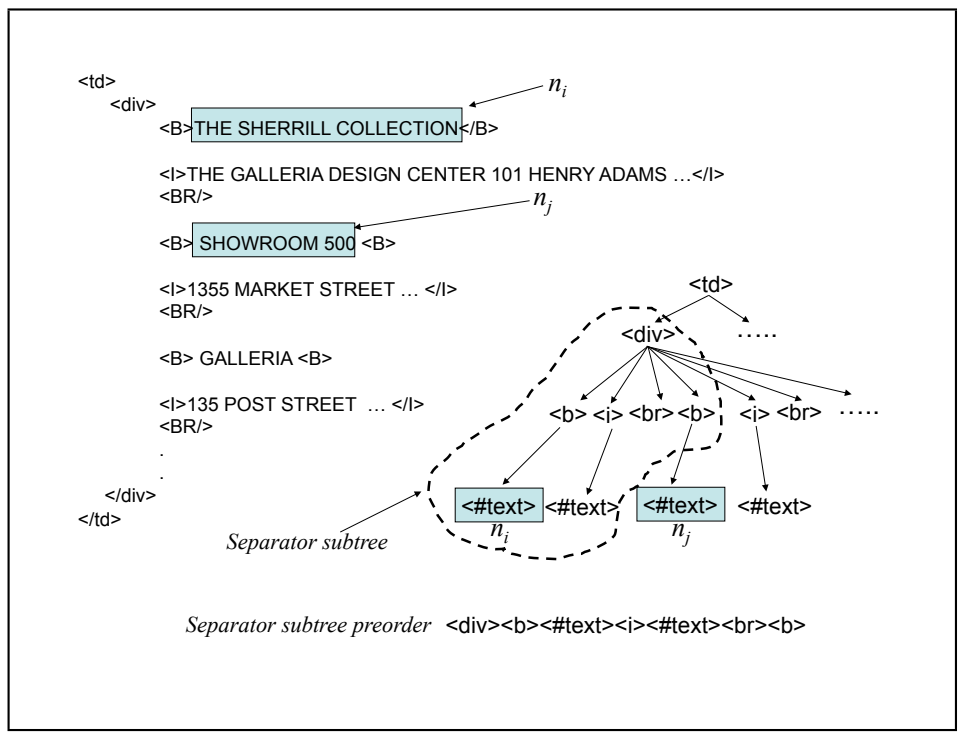

**Figure 7: Boundaries of records spanned by a given wrapper.**

| | |
|---|---|
| 1. | `cduniverse.com` |
| 2. | `music.barnesandnoble.com` |
| 3. | `tower.com` |
| 4. | `cdbaby.com` |
| 5. | `musicishere.com` |
| 6. | `home.napster.com` |
| 7. | `mog.com` |
| 8. | `mp3.rhapsody.com` |
| 9. | `shockhound.com` |
| 10. | `rollingstone.com` |
| 11. | `play.com` |
| 12. | `wayango.com` |
| 13. | `audiolunchbox.com` |
| 14. | `amazon.com` |
| 15. | `allmusic.com` |

**Figure 8: List of websites used in DISC dataset.**

| | |
|---|---|
| 1. | *Bach for Breakfast*, Johann Sebastian Bach. |
| 2. | *Abbey Road*, Beatles. |
| 3. | *If It Rains on Tuesday*, Michelle Suesens. |
| 4. | *Notre Dame Lullabies*, The O'Neill Brothers. |
| 5. | *Love is the Answer*, Barbra Streisand. |
| 6. | *Strangers In the Night*, Frank Sinatra. |
| 7. | *I Left My Heart In San Francisco*, Tony Bennett. |
| 8. | *Au Nom d'Une Femme*, Helcne Segara. |
| 9. | *Yesterday & Forever*, Beatles. |
| 10. | *Mi Plan*, Nelly Furtado. |
| 11. | *She Walks In Beauty*, Danielle Woerner. |

**Figure 9: List of albums used in DISC dataset.**

For instance, it is hard to imagine a CRF that can extract business names from a collection of webpages, and do it over any given store website with any reasonable accuracy. An interesting piece of work [20], with the exact problem setting as ours, use (an extension of) CRFs for information extraction, and learn them automatically from noisy labels. The reported accuracy, however, ranges between 60% to 80% based on 10 websites, which shows the fundamental limitation of this approach. Our experiments cover more than 300 websites yet achieve close to 100% accuracy. Ensemble-learning techniques have been applied to the problem of schema matching with noise [12]. These techniques, however, are bagging-type: they assume that the underlying schema matcher is already noise-tolerant to a certain extent. Wrapper inductors, on the other hand, are very sensitive to even small amounts of noise.

The second line of work, which is more directly related to ours, is that on grammar induction techniques [5, 21, 2], which learn a grammar to capture the repeating structure of pages in websites in a completely unsupervised way. Various languages (e.g., union-free regular expressions) have been proposed as good abstraction for describing the structure of websites, and these works try to find the minimal grammar in these languages that generate the given set of webpages. Note however that a learned grammar does not directly translate into an extraction rule. For example, consider a table where the second `td` in each `tr` is the business name that we want to extract. A grammar inductor can easily infer the grammar for a table as $(\texttt{tr}(\texttt{td}^*))^*$. However, the underlying problem still remains: figuring out the rule that the second `td` in each `tr` is the business name. In general, a grammar is much more complex, and by itself, it does not tell how to extract the types that are of interest to us. In this work, we use ideas from grammar induction, and combine them with supervised wrapper induction methods.

Philosophically, our approach to scalable information extraction can be deemed as *domain-centric*, in contrast to some of the recent approaches for scaling extraction, which are *web-centric*. The *WebTables* [4] and its extension to list extraction [7] and the NLP-based *KnowItAll* system of Etzioni et al. [8] are well-known web-centric approaches. By being domain-centric, (1) we are able to leverage domain-knowledge, dictionaries for annotations, and accurate models for identifying lists in the domain, and (2) our extractors (by definition) also attach types to the extracted data, in contrast to web-centric approaches, that need a separate schema matching process to attach semantic meaning to the extraction results: a challenging task on a web-scale. Our approach is the first to achieve very high accuracy in scaling domain-specific extraction to the entire web.

**Ensemble-based methods.** Ensemble-based methods use multiple models to obtain better predictive performance than individual constituent models, and have been successfully used in Machine Learning to handle noise in input labels. However, these techniques do not apply well to the wrapper learning problem. We describe here some of the ensemble techniques and their limitations.

*Multi-strategy learning* [9] assumes that there are multiple learners making "complementary" errors when given noisy labels, and combine the predictions of multiple matchers using some sort of a voting scheme to obtain accuracy higher than that of any individual learner. However, the classic wrapper learning algorithms only work on a clean set of annotations, and when given noisy annotations, perform really bad and moreover, make the same errors. E.g. every learner will try to fit all the input annotations, and every incorrect annotation will be in the output of all the learners. Thus, combining the predictions will not generally reduce the noise.

Further, multi-strategy learning is a supervised technique. In our setting, we cannot evaluate if a wrapper rule is correct, since our input annotations are not only noisy, but also incomplete (they only match a subset of items on each page).

Another closely related ensemble technique is called *bagging* [3]. In bagging, a single learner is trained on multiple randomly-drawn subsets of the training data, and the resulting set of models are combined using some ensemble scheme. This method has been applied for problems like schema matching with noisy attributes [12]. The idea is that each sample contains fewer errors, and further, different models are learnt on different sets of errors. So, when they are combined, errors are canceled out. However, again, bagging works when the underlying learner has at least some tolerance to noise. In our setting, even one error in annotation results in a highly over-generalized wrapper. Thus, wrappers are not very amenable to bagging-style approaches.